# Optimized nanodevice fabrication using clean transfer of graphene by polymer mixture: Experiments and Neural Network based simulations

Jared K. Averitt, Sajedeh Pourianejad, Olubunmi Ayodele, Kirby Schmidt, Anthony Trofe, Joseph Starobin*, and Tetyana Ignatova*

*Department of Nanoscience, University of North Carolina at Greensboro, Greensboro, North Carolina 27401, United States*

JKA, SP contributed equally

**Abstract**

In this study, we investigate both experimentally and computationally the molecular interactions of two distinct polymers with graphene. Our experimental findings indicate that the use of a polymer mixture reduces the transfer induced doping and strain in fabricated graphene devices as compared to conventional single polymer wet transfer. We found that such reduction is related to the decreased affinity of mixture of polymethyl methacrylate and angelica lactone polymer for graphene. We investigated changes in binding energy (BE) of polymer mixture and graphene by considering energy decomposition analysis using a pre-trained potential neural network. It was found that numerical simulations accurately predicted two-fold reduction of BE and order of magnitude reduction of electrostatic interaction between polymers.

## Introduction

Among 2D materials, graphene has been the most extensively studied due to remarkable properties [1-4] and promising applications. The chemical vapor deposition (CVD) method remains the most reliable to produce high quality and large area graphene on Cu [5,6], Ni [7] or Pt [8,9] surfaces. Even though the CVD-grown graphene usually consists of a single or sometimes multiple layers of graphene, [10-12] it is unusable on the growth metallic surface, and thus clean transfer on to target substrate such as Si/SiO$_2$ [13], glass [14], polyethylene terephthalate [15] or paper [16] is crucially important for various applications ranging from biomedical [17] to nanoelectronics and quantum computing [18]. Several transfer methods are in use [19-25], but the polymer support method is a promising one because it could be easily scale up for industrial implementations [26, 27]. The polymer of interest includes polycarbonate [28], polydimethylsiloxane, [7] but the most popular one is polymethylmethacrylate (PMMA) [29,30]. However, inconsistency in the quality of graphene transferred with PMMA has limited its application in device fabrication. This inconsistency is attributed to the presence of carbonyl functional groups (C=O) [31] and long chain structures [32] which are contributing to high binding energy of PMMA to graphene and cause incomplete removal from 2D surface after transfer to device substrate (here we are not discussing graphene imperfections such as defects, grain boundaries, edges, etc. [27]) Additional aggressive solvent treatment (either hot or fuming acetone) [24, 33] or thermal annealing [34] did not significantly improve PMMA removal. Other cleaning methods based on either UV/ozone treatment [35] or argon beam bombardment [36] have been employed but cause graphene quality reduction. There are reports of other less aggressive alternative methods requiring complicated equipment setups and involving the use of two layers of PMMA [34, 37] which further cause appearance of additional wrinkles and cracks in graphene during transfer [35]. The efficiency of graphene transfer can be improved by blending PMMA with a polymer having a low binding energy to graphene [36, 37]. In this work we demonstrated large area, clean graphene transfer using PMMA and an additive, the polyfuranone chain products produced from biomass-derived angelica lactone via C-C coupling reaction, which we will call ALP for simplicity (Fig. 1 c, d) [38]. Understanding the physical mechanisms behind binding polymer molecules on graphene is a challenging computational problem. Indeed, the binding cannot be described as a single global minimum of a potential energy since polymer molecules are not covalently attached to graphene surface. To address this challenge, we used a potential neural network-based approach to calculate minimal energy configurations of graphene and polymer mixture by considering multiple initial conditions for positions of polymer atoms (high throughput cycle as shown in Fig. 2). This method was chosen to circumvent time related deficiency of electron configuration calculations typical for DFT based simulations.

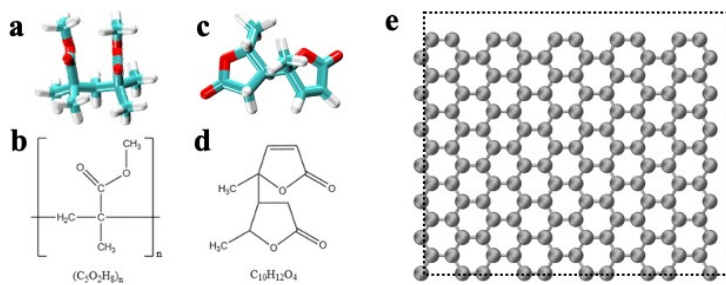

*Figure 1: The gas phase relaxed molecules. PMMA (a) and (c) ALP polymer. Structural formulas of PMMA (b), ALP (d). 5x7 unit cell of graphene was used as the graphene model top view (e) with periodic boundary conditions shown in dashed line.*

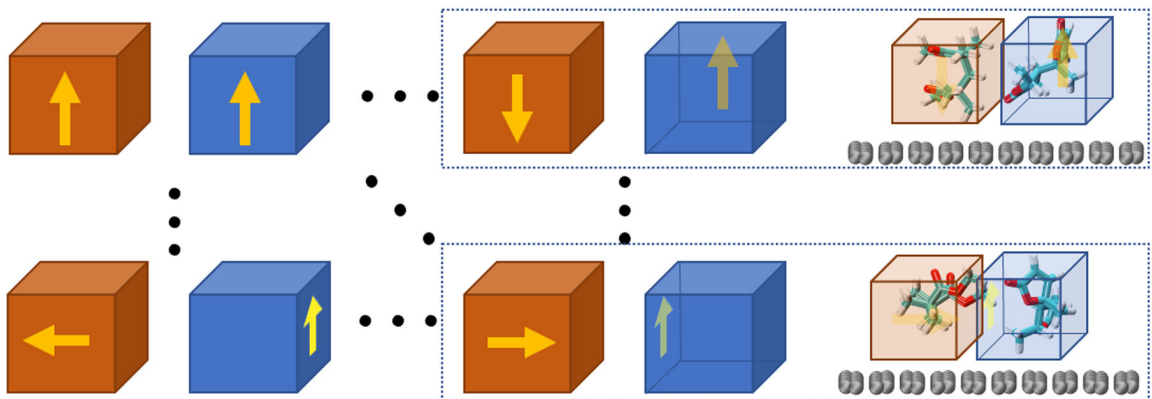

*Figure 2. Schematic representation of the molecular rotation algorithm used to generate confirmations. The orange cube represents one polymer molecule while the blue represents the second polymer molecule.*

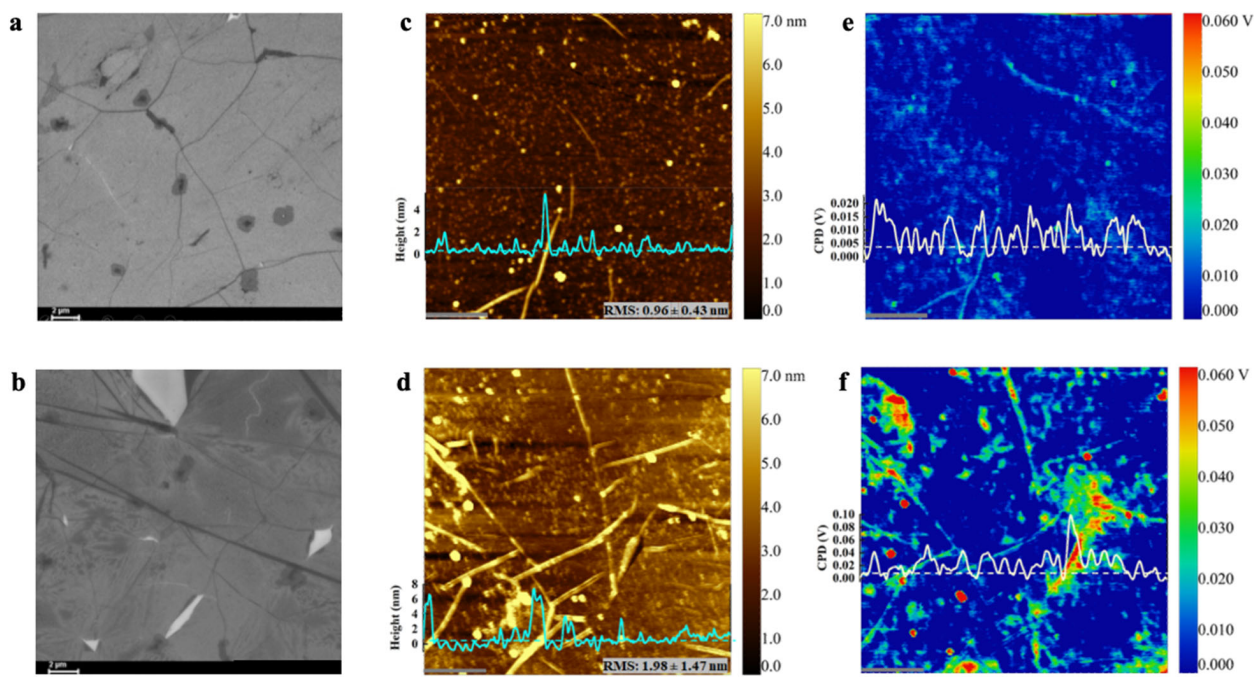

*Figure 3: the SEM images of graphene transferred with (a) ALP:PMMA and (b) PMMA; area of scan is 135 μm2; the dark spots observed on both images are atmospheric water molecules adsorbed on the surfaces of graphene. AFM maps of graphene transferred with (c) ALP:PMMA (RMS =0.96 ± 0.43 nm) and (d) PMMA (RMS =1.98 ± 0.47 nm); the inserts show the line profiles and surface roughness. KPFM maps of graphene transferred with (e) ALP:PMMA and (f) PMMA and line profile of surface contact potential. Scale bars are 1 μm.*

**Experimental Approach and Results of Experiments**

Figure S1 shows the procedure of the proposed process for transferring CVD graphene. In the conventional transfer method, PMMA is typically spin-coated on the graphene-on-growth substrate. In this work we mixed the solutions of PMMA and ALP at different weight concentration ratios of ALP:PMMA, as [1:1], [1:2], [1:4], [1:6], [1:0] and then spin-coated on CVD graphene grown on Cu foil. It has to be noted that ALP stays as a jelly like substance after all solvent removal, even after cooling down polymer to $4^0$ C, therefore we could not use ALP alone as a sacrificial layer in transfer procedure. The rotation speed was adjusted to get thickness of polymer film approximately 1 μm. After spin-coating, samples were dried at room temperature for 24 h and then soft-baked at temperature 95°C for 5 min to evaporate solvent. Cu foil was delaminated by applying the "bubbling procedure" which is basically a water electrolysis process, in details described previously [39, 40]. We observed that the polymer graphene stack was detached from the Cu foil very effectively and fast (3-5 seconds) for ALP:PMMA concentration of [1:4], leaving behind clean grown substrate. After cleaning with de-ionized water, the floating polymer-graphene "sandwich" was deposited on Si/SiO$_2$ substrate and dried gradually at 90 – 135°C for 30 min. Finally, the sacrificial layer made of polymer mixture was removed by acetone in Soxhlet extractor to prevent any contamination from solvent side [20]. We applied multiple characterization techniques to compare quality of transferred material. Scanning electron microscopy (SEM) images of graphene transferred using ALP:PMMA [1:4] showed fewer defects and polymer residues (Fig. 3 a, b) in comparison to graphene transferred with PMMA only. Results for other concentrations of ALP:PMMA could be found in SI. The concentration of polymer mixture ALP:PMMA [1:4] will be used for all future considerations and will be compared to the PMMA-only transferred graphene. To analyze polymer residues on graphene surface, we used atomic force microscopy (AFM). With great consistency to previous reports [41-43] the PMMA transferred sample showed the presence of a few intermittent cracks in graphene and moderate polymer residues (RMS=1.98 ± 0.47 nm). The higher quality of the ALP:PMMA transferred sample (Fig. 3 c, see line profile in insert) with RMS= 0.96 ± 0.43 nm could be attributed to (1) softening of polymer blend by adding jelly-like ALP, allowing to evenly distribute introduced by polymer strain and so minimizing appearance of graphene cracks; (2) decreasing of adhesion of polymer blend vs. PMMA, resulting in easier removal sacrificial layer from graphene surface during the final transfer step. Of course, we cannot prevent attachment of polymer residues at defected sites of graphene which in turn would result in the formation of strong covalent bonding between graphene and polymer molecules as shown by Leong et. al. [31]. The proposed novel ALP:PMMA transfer method significantly reduces graphene damage, thus only non-covalent interactions between graphene and polymer sacrificial film are expected and numerical calculations will provide more details on these. The adsorbed polymer residues can significantly reduce charge carrier mobility of graphene [43-45]. Therefore, the transport properties of graphene can be improved by minimizing the polymer residues [41-43,47-49]. The Kelvin Probe Force Microscopy (KPFM) characterization reveals homogeneous surface potential distribution over large area of graphene transferred with ALP:PMMA (Fig. 3 e and insert). Whereas twice higher variations in the surface potential distribution were observed in the PMMA-transferred graphene sample (Fig. 3 f and insert) reflecting introduced parasitic graphene doping (Fermi level shift). Kim et.al., reported that the origin of this inhomogeneity is directly related to PMMA residues [50].

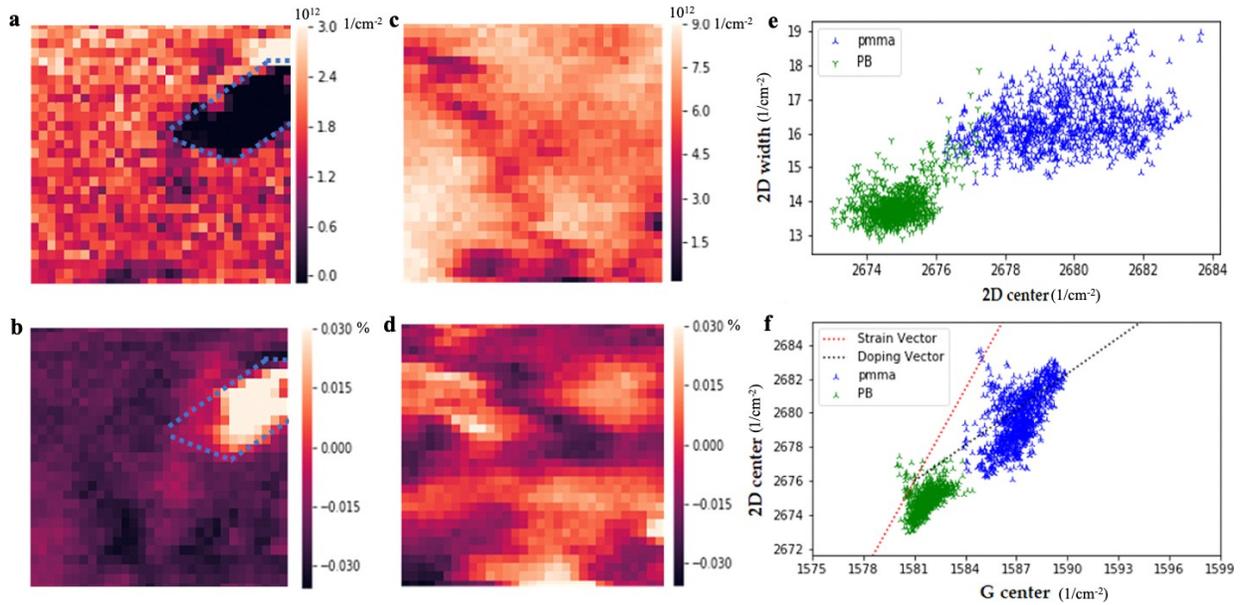

*Figure 4: Analysis of the quality of graphene using Raman Spectroscopy. Doping (a) and strain (b) map of graphene transferred by polymer blend method; blue dotted region is few layers of graphene. Doping (c) and strain (d) map of graphene transferred by PMMA method. The correlation plot of 2D band center vs width (e). 2D band center vs G band center (f). scale for(a)-(d) is 0.5µm per pixel.*

Hyperspectral Raman characterization is known to be a powerful tool to examine quality, layer number as well as quantify local doping and strain variations over large area of graphene [51-53]. We performed Raman mapping for samples transferred by polymer blend and PMMA. The spectra were fit using least squares minimization of Lorentzian peaks. The position, broadening and shift of the Raman characteristic peaks of graphene (D, G, and 2D) were analyzed. From the correlation plots in the Figure 4 e, f, we clearly see that the PMMA-transferred sample has a wider 2D peak indicating reduce charge carrier doping, and a significant shift in both the G and 2D peaks when compared with ALP:PMMA-transferred sample. In accordance with [51], strain and doping induce shifts in the characteristic peaks (2D and G) of graphene. The G peak is particularly responsive to doping, while the 2D peak is influenced by strain. Plotting the 2D against the G peak provides a visual representation of the level of strain and doping. Points at the intersection indicate zero strain and doping. As strain is applied, peak positions shift along the red curve (Fig.4 f), with both G and 2D peaks moving to lower wave numbers for tensile strain and higher for compressive strain. Increased p-doping shifts the peak band along the magenta curve. It's important to note that this procedure is only applicable to monolayer graphene, hence a few-layer graphene area in the ALP:PMMA-transferred sample was excluded as an outlier from the scatter plots. The corresponding doping maps in Figure 4 a, c strongly aligns with the KPFM results, showing a substantial reduction in parasitic p-doping in the ALP:PMMA-transferred graphene compared to PMMA-transferred graphene. In Figure 4 a, c, green hexagons denote multilayer graphene areas. Additionally, in Figure 4 c, d, we observe both high and low strain areas in both samples, with the ALP:PMMA-transferred sample exhibiting a more uniformly distributed strain.

**Results of Numerical Simulations**

**Introduction to numerical methodology**

As mentioned above we have developed a new approach to perform high throughput cycle atomic simulations to quantify nonlocal interactions on 2D interface between graphene and non-covalently bonded molecules. To create a matrix of atomic parameters (positions, masses, energies and forces) we used an atomic simulation environment (ASE) [54] (Fig. 5 a). At each step of matrix update the atomic level forces have been determined using the potential neural network (PNN), ANI-1ccx [55]. After that the positions were updated to determine dynamic parameters of nuclei by using the Broyden-Fletcher-Goldfarb-Shanno (BFGS) optimization predictor-corrector numerical process (Fig. 5 b) [56]. The atomic charges $q_i$ of each molecule were calculated separately with respective atomic parameters and relaxed geometries [54] using the full Geometry-dependent Atomic Charges (GDAC) [57] method (Fig 5 c). We utilized Multiwft [58] software package to perform energy decomposition analysis using classical force field (EDA-FF) similar to that described in [59]. The binding energy (BE) of each system (Fig. 1) was normalized per unit area of van der Waals overlap. The van der Waals area of overlap is the region where the binding energy can be approximated by a Lennard-Jones potential. This region is where the polymer atoms are in close proximity of the 2D interface. The values we used for the van der Waals radii: C, H, O atoms are 1.77, 1.2, 1.52 A respectively [58].

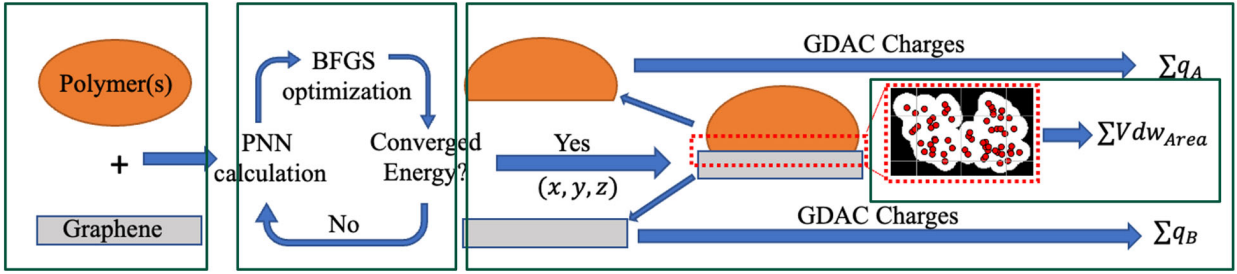

*Figure 5. Schematic representation of the algorithm used. PNN/BFGS iterates until energy convergence of 0.05 meV. Outputs of atomic positions (x, y, z), charges (q) and the van der Waals area are computed from the optimized structures.*

**High throughput cycle description**

The matrix of atomic parameters (which will be called matrix from here on) is initialized before going through the PNN (Fig. 5). We initialized the initial positions using a series of rotations relative to a single input configuration (Fig 2 a). The rotations are performed using Euler rotations, where phi, theta, and psi are the Euler angles, and the molecules center of masses are the center of rotation. In order to generate 576 configurations of equal-proportion displacements (Fig. 2), we generated 6 sets of conformations (representing each face of a 6-sided cube), with each set consisting of 4 conformations (representing $90^0$ rotations along each face of the cube).

**Atomic charge calculations**

To perform EDA-FF calculations one needs to determine the charge of each atom. The PNN does not provide these charges, so we implemented a geometry dependent atomic charge (GDAC) method [60].

**Energy Decomposition Analysis using Force Fields (EDA-FF)**

The converged matrix positions and the charges from the cycle (Fig 5 c) are used to calculate the binding energy (BE) through Energy Decomposition Analysis using Force Fields (EDA-FF). By calculating the binding energies ($\Delta E$) of the respective polymer-graphene interacting system $\sum E(C_i)$ and energies after disassociation $E(C_1, C_2, \ldots, C_N)$, a numerical comparison of the adhesion ability of the polymer(s) and graphene are obtained. The binding energy between N molecular fragments:

$$\Delta E = E(C_1, C_2, \ldots, C_N) - \sum E(C_i) \quad (1)$$

Strongest adhesion is directly proportional to larger negative values of the BE. Energy Decomposition Analysis using Force Fields (EDA-FF) is an attractive method due to its requirement of only optimized structures and atomic charges as inputs. This feature makes it computationally efficient, requiring negligible resources (< 5 seconds per calculation on a single CPU). EDA-FF calculates BE as three separate terms: electrostatic ($\Delta E_{es}$), short-range (exchange) repulsion ($\Delta E_{ex}$) and long- range dispersion ($\Delta E_{disp}$).

$$\Delta E = \Delta E_{es} + \Delta E_{ex} + \Delta E_{disp} \quad (2)$$

where the electrostatic energy (Coulomb potential) between atoms A and B is:

$$E_{es} = \frac{q_A q_B}{r_{AB}}$$

and the van der Waals interaction energy (Lennard Jones potential) between atoms A and B is the sum of the repulsive interaction due to Pauli repulsion:

$$E_{ex} = \varepsilon_{AB} \left(\frac{R^0_{AB}}{r_{AB}}\right)^{12}$$

and the attractive dispersive interaction (dispersion).

$$E_{disp} = -2\varepsilon_{AB} \left(\frac{R^0_{AB}}{r_{AB}}\right)^{6}$$

The $\varepsilon_{AB}$ is the well-depth of interatomic van der Waals interaction potential, while $R^0_{AB}$ is the van der Waals radius. The $r_{AB}$ is the distance between atom A and atom B. The interatomic parameters $\varepsilon_{AB}$ and $R^0_{AB}$ are provided by the trained force fields and the values are commonly defined for each atom type:

$$\varepsilon_{AB} = \sqrt{\varepsilon_A \varepsilon_B} \quad , \quad R^0_{AB} = R^*_A + R^*_B$$

where $\varepsilon_{AB}$ and $R^*_B$ are parameters defined by the AMBER atom types which are available here [46].

**Van der Waals sphere half area of overlap**

Using information about the van der Waals radii and relative positions of each atom in a system we are able to calculate the overlapping van der Waals area between the polymer(s) and the graphene surface. First, we listed the graphene atoms and grouped them with the PyVista [60] spheres that represent them. Next, we checked for any overlapping spheres between the graphene and polymer atoms by comparing their distances and van der Waals radii. If we found any overlapping spheres, we recorded the indices of those atoms. We then merged all the overlapping graphene atoms and polymer atoms together into a single mesh object. We took the boolean intersection of these two new object meshes, and calculated the area of that intersection. This method allowed us to avoid overcounting the area in cases where two or more van der Waals spheres overlapped with a single atom. Finally, we take a factor of 1/2 to avoid double counting the area.

**Constraints over configurational space calculations**

To obtain meaningful results when exploring a large range of configurations, constraints are necessary. We excluded configurations that did not meet our selection criteria, which included a van der Waals sphere half area of overlap greater than $2\text{Å}^2$ and a negative binding energy. Only the systems that satisfied these criteria were considered for analysis, and the minima energies were reported in Table 1. We chose this selection criteria because a positive binding energy does not have physical meaning, while a van der Waals sphere half area of overlap less than $2\text{Å}^2$ suggests that the molecules are too far apart to interact non-locally, resulting in an unphysical system.

**Comparison of polymer-polymer and polymer(s)-graphene interactions**

Quantification of the interactions at the polymer-graphene interface is represented by binding energy per unit area of van der Waals overlap. We employ a gaussian fit over all possible energies of the relaxed geometries within the constraints (Fig. 6). The labeling convention for each model in the computational results corresponds to ALP, PMMA and graphene as *A, P* and *G* respectively. Furthermore, the subscripts G, A, P correspond to the $C_0$, $C_1$ values that correspond to *G, A* or *P* of each model used in the calculation of eq. 1. Table 1 shows the results of different models and their corresponding contributions to the binding energy values per unit area ($\Delta E$). The models include $GAP_{GP}$ (i.e. interaction of GP for model *GAP*), $GAP_{AP}$, *GP* (i.e. interaction of GP for model *GP*), *AP, PP*, and *AA*. All the results have negative total energy values, indicating that the systems are stable. The largest negative total energy value is −2.5 [meV/particle/Å$^2$] in the *GP* model, while the smallest is −1.2 [meV/particle/Å$^2$] in the $GAP_{GP}$ model/dimmer. This significant reduction in the binding energy per unit area is indicative to the mix of polymers having a strong decrease in their binding energy with graphene when compared to just PMMA alone.

*Table1. Minima values from EDA-FF corrected with van der Waals sphere half area of overlap, physical selection criteria: $E_{int}$<0 and vdW area > 2Å$^2$*

| Model$_{(dimer)}$ | Energy/Area [meV/particle/Å$^2$] | | | |
| --- | --- | --- | --- | --- |
| | (total) $\Delta E_{int}$ | (attractive) $\Delta E_{es}$ | $\Delta E_{disp}$ | (repulsive) $\Delta E_{ex}$ |
| $GAP_{(GP)}$ | -1.2 | -0.1 | -1.1 | -0.1 |
| $GP$ | -2.5 | -0.1 | -3.0 | 0.5 |
| $GAP_{(AP)}$ | -2.2 | -0.1 | -5.2 | 2.9 |
| $AA$ | -1.7 | -0.2 | -2.3 | 0.4 |
| $AP$ | -2.3 | -0.1 | -5.0 | 2.6 |
| $PP$ | -1.5 | 0.01 | -2.4 | 0.9 |

In terms of the energy components, electrostatic and dispersion energies are always negative, while repulsion energies are positive, which is true for all stable systems.

Comparing the models, the $GAP_{GP}$ model has the lowest repulsion energy, while the $AP$ model has the highest. The $GAP_{AP}$ model has the highest dispersion energy, while the $GAP_{GP}$ model has the lowest. Overall, the table provides a useful summary of the energy values for different models and can help guide further analysis and understanding of the systems being studied. PMMA has a significant reduction in BE (1.3 meV) when both polymers are on graphene (i.e. $GAP_{GP}$). This is consistent to the experimental observations of lower residue concentration on the polymer blend used for graphene transfer. This table provides a more in-depth picture to the successful transfer we observed and the less amount of residues and uniform properties observed.

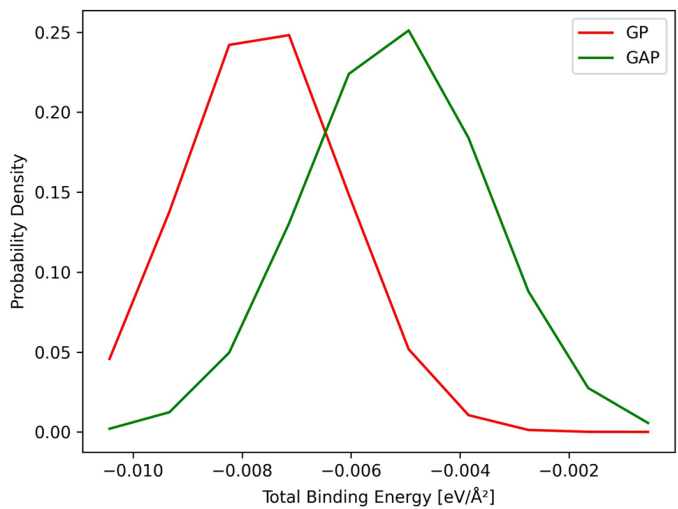

*Figure 6. Gaussian fit of binding energies [eV /Å$^2$] for the models: GP and GAP$_{GP}$ (labeled as GAP here).*

In the simulations of the graphene surface, a periodic array consisting of 5 × 7 graphene unit cells was utilized, with all carbon atoms of the graphene being constrained in all directions. We simulated PMMA as a fragment with $n = 2$ and ALP as a dimmer of 2 lactone rings.

## Conclusions

We demonstrated clean and large area graphene transfer by using polymer blend with optimized ratio of two polymers. In addition, we designed an algorithm for numerical calculations follows the experimental trend and provides a novel and effective approach for quantifying non-local interactions in multi-molecular systems. This suggests that considering the van der Waals sphere half area of overlap reveals a better representation of the underlying physical interactions in molecular systems on the surface of graphene. The approach allows for reporting energy units to energy/area, which is consistent with the trend in experimental results.

## Instrumentation

The samples 'morphologies were obtained with a scanning electron microscope (Zeiss Auriga FIB/FESEM, Jena, Germany) at an accelerating voltage of 5 kV. The surface roughness was obtained using Oxford Research AFM (MFP-3D infinity, Santa Barbara, Ca, USA) in the tapping mode at ambient conditions. A Si tips coated with Al (TAP300AL-G probe, Budget Sensors) was used for the topological probing. The amplitude modulation mode in Kelvin probe force microscopy (AM-KPFM) was employed for the measurement of contact potential difference (CPD) of the transferred graphene. A conductive probe consisting of Pt/Ir-coated tip (EFM, Nanoword) was used while silver paint served as the ground. Raman spectra were measured using a Horiba XploRa Raman Confocal system (Kyoto, Japan) with an excitation wavelength of 532 nm and a 1200 L mm-1 diffraction grating. The mapping of the total coverage of graphene (4 μm x 4 μm) resulting in 2000 data points were collected using a 100x objective in x-y-z directions.

## Acknowledgements

J.K.A. acknowledges that this material is based upon work supported by the National Science Foundation Graduate Research Fellowship under Grant No. [1945980]. T.I., J.K.A., A.T. acknowledge the US Department of Defense [Contract #W911QY2220006]. This work was performed at the Joint School of Nanoscience and Nanoengineering, a member of the Southeastern Nanotechnology Infrastructure Corridor (SENIC) and National Nanotechnology Coordinated Infrastructure (NNCI), which is supported by the National Science Foundation [ECCS-1542174]. T.I., J.K.A., A.T. acknowledge the 2DCC grant as NSF cooperative agreement DMR-1539916.

**Optimized nanodevice fabrication using clean transfer of graphene by polymer mixture: Experiments and Neural Network based simulations.**

Jared K. Averitt, Sajedeh Pourianejad, Olubunmi Ayodele, Kirby Schmidt, Anthony Trofe, Joseph Starobin*, and Tetyana Ignatova*

Nanoscience Department, University of North Carolina, Greensboro, United States of America.

JA and SP contributed equally.

## 1. Schematic of transfer process

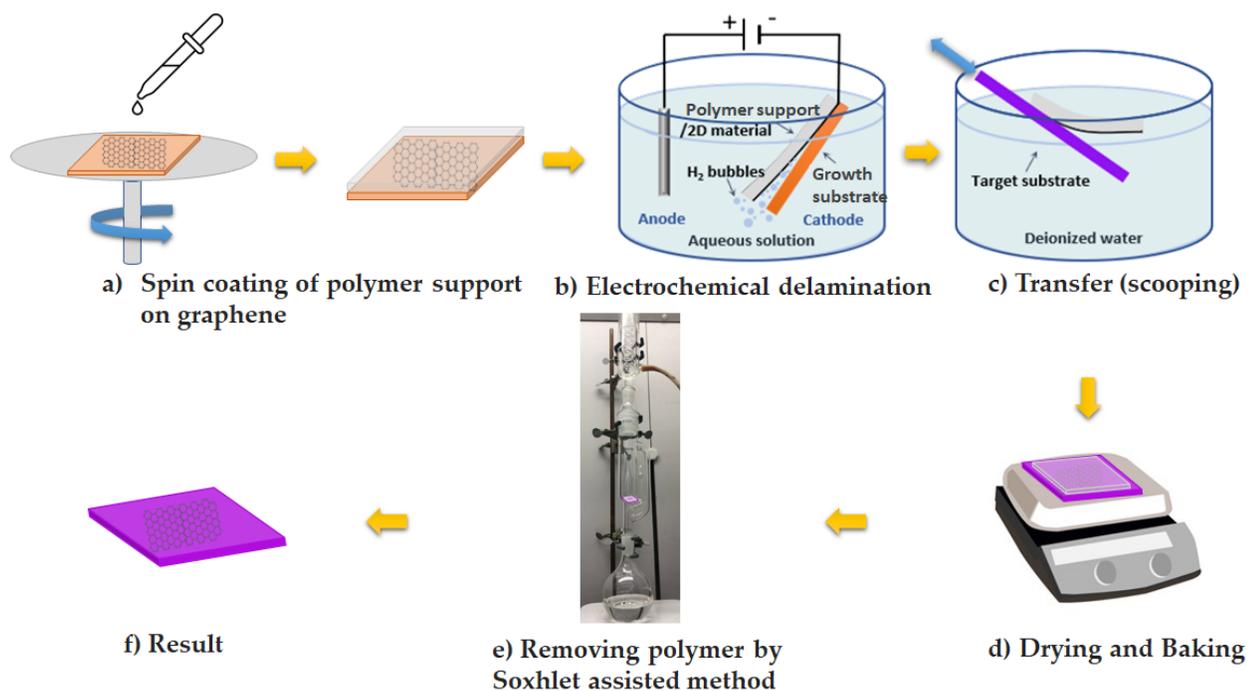

Figure S1. Schematic diagram of graphene transfer process. This scheme was also used for PMMA-assisted transfer. In a bid to reduce transfer error, multiple transfers were carried out and the variability between each transfer was noted.

## 2. Raman spectrum

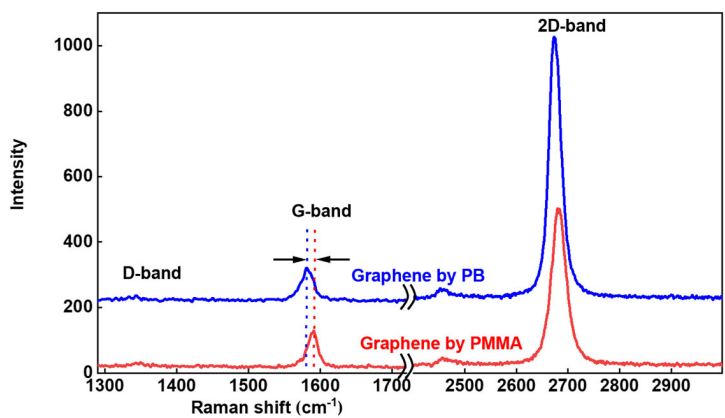

Figure S2. Raman spectra of graphene transferred by polymer blend (blue) and PMMA (red).

## 3. Optimizing the appropriate ratio of ALP and PMMA

Polymer preparation and characterization of polyfuranone chain products: The polyfuranone chain products (PCP) were derived from the reaction of AL at 80°C for 5 min. Under K2CO3 catalysis, the reaction was spontaneous thus achieving complete conversion to its corresponding polyfuranone - dimers (64 %), trimers (34 %) and trace quantity of tetramers (2 %) (Xin et al., 2014; Ayodele et al., 2017). The product was purified by DI water treatment. The functional groups were determined by Fourier transform infrared spectroscopy (Agilent 670 FTIR Spectrometer w/ATR (Santa Clara, CA, USA) (SI). The glass transition state of the PCPs was obtained by differential scanning calorimetry (Perkin Elmer DSC6000, Waltham, MA, USA). The measurement was obtained by heating the sample from 0 to 150°C (heating rate of 10°C/min) in an inert atmosphere (SI).

Blending of polymers has been a favorable practice in the polymer industries but rarely used in graphene transfer processes, and in most cases, it is used to confer a certain physical/chemical property on a polymer. In graphene transfer, PMMA is commonly used, but due to its strong interaction with graphene [1], clean transfer is rarely achieved [2]. In this work, we altered the strong chemical interaction between graphene and PMMA with a low molecular weight polymer, ALP, and an appropriate blending ratio to yield a clean graphene was investigated with AFM and SEM.

As seen in Table S1, the mean roughness (RMS) calculated from AFM characterization of the PMMA-transferred graphene was higher than the RMS value of the polymer blend transferred sample. Results are provided for the ALP:PMMA ratio of 1:4 – 1:6 having the lowest RMS of 0.96 and 0.748 nm, respectively. The SEM images are shown in Fig. S5 and they confirmed that ratios 1:4 – 1:6 have fewer defective features. Factors that contributed to their improved surface roughness and morphology could be the enhancement in support strength of the blended PMMA thus preventing wrinkles or tear of graphene during transfer processes. In other to understand the chemistries of the polymer adducts, we studied the glass transition temperature ($T_g$) of the material (ALP:PMMA = 1:4) in respect to the standalone ALP and PMMA (Supplementary Fig. S6). Despite the high miscibility of the two polymers, a physical blending at room temperature (RT) does not results in the formation of new type of polymer and further proof was obtained by non-appearance of new functional groups (Fig. S7). However, the shift in absorption bands at line a (C=O stretching mode) and the disappearance of line b in ALP (-OH stretching due to $H_2O$ physisorption) further confirmed a change in polymer geometry. Based on the calorimetric measurement, the $T_g$ of the polymer blend is within the $T_g$ of PMMA irrespective of the mixing ratios indicating that the two polymers do not bind strongly together.

Table S1. Comparative analysis of the surface roughness of graphene obtained using the blended polymers

| Sample | RMS (nm) |
| --- | --- |
| PMMA | 2.69 ± 0.80 |
| ALP:PMMA 1:6 | 0.748 ± 0.76 |
| ALP:PMMA 1:4 | 0.96 ±0.61 |
| ALP:PMMA 1:2 | 1.532 ± 1.07 |
| ALP:PMMA 1:1 | NA |
| Si before transfer | 0.13 ± 0.02 |
| Si after transfer | 0.6 ± 1.34 |

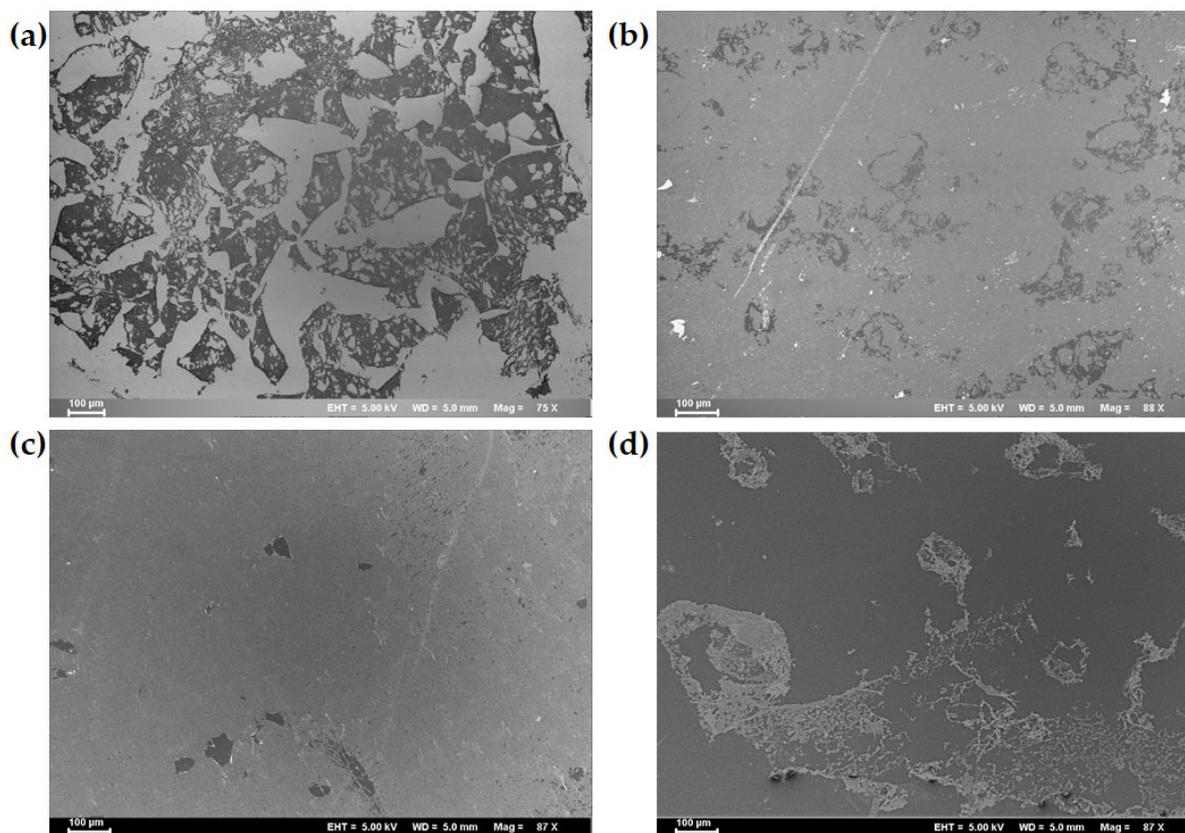

Figure S3. SEM images of graphene transferred by PMMA(a), (b) ALP:PMMA 1:6, (c) ALP:PMMA 1:4, (d) ALP:PMMA 1:2.

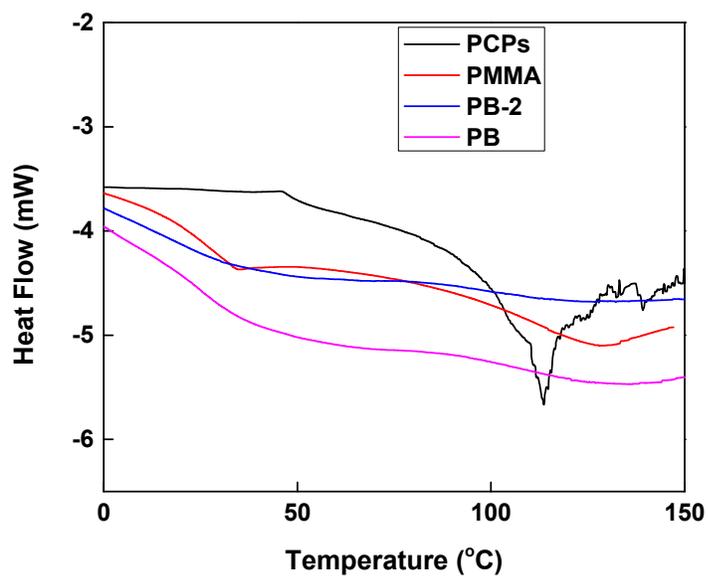

Figure S4. DSC curves of ALP (black), PMMA (red), ALP:PMMA 1:2 (purple) and ALP:PMMA 1:4 (blue)

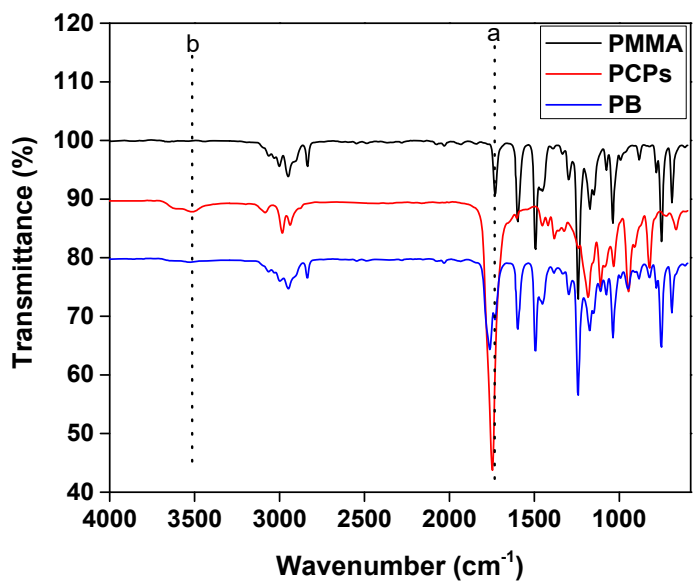

Figure S5. FT-IR of ALP (red), PMMA (black) and ALP:PMMA 1:4 (blue)

## 4. Surface energy and surface tension calculation:

To estimate the binding energy experimentally we performed contact angle measurements, where surface energy values were calculated for each polymer, polymer mixture and graphene on Si/SiO2 (The graphene partial transparency theory implies that graphene surface energy is dependent on supporting substrate, Si/SiO2 in our case). To measure free surface energy of graphene, we used equation (1), resulted from the Girifalco-Good-Fowke's Young equation [3,4]:

$$\gamma_G = \frac{(\gamma_{H_2O}(1+\cos\theta))^2}{4\gamma_{H_2O}^d} \tag{1}$$

Where, $\gamma_{H_2O}$ is the surface tension of the water drop, $\gamma_G$ is the free surface energy of graphene (solid surface), $\gamma_{H_2O}^d$ is water dispersive interactions, and $\cos\theta$ is the contact angle between the liquid-vapor interface and the solid surface. The relation between interfacial tension of solid surface and the solid–liquid interface can determine whether contact angle ($\theta$) is either less or greater than 90°, which is an interpretation of the wettability of the surface. If $0 < \theta < 90°$, the liquid partially wets the solid and the surface is said to be hydrophilic. The hydrophobicity rises as the contact angle of the droplets with the surface increases. Hence, hydrophobic surfaces have contact angles larger than 90°.

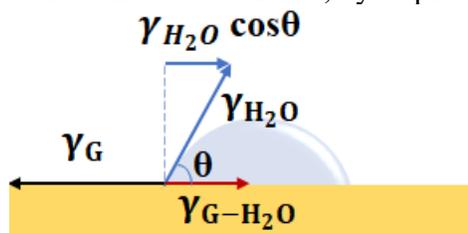

Figure S6. Contact angle schematic.

The surface tension of polymer can be estimated by the molar parachor, which was introduced by Sugden (1924), who defined a list of atom-groups' contributions [5];

$$\gamma_P = \left(\frac{P_s}{V}\right)^4 = \left(\frac{P_s \times \rho}{M}\right)^4 \tag{2}$$

$\gamma$ (ALP) = 13.22 mJ/ m²
$\gamma$ (PMMA) = 42.5 mJ/ m²
For ALP:PMMA 1:4    $\gamma$ (polymer blend) = (1/5) × 13.22 + (4/5) × 42.5 = 36.64 mJ/ m²
For ALP:PMMA 1:2    $\gamma$ (polymer blend) =  32.74 mJ/m²

where $\gamma_P$ is surface tension of polymer, Ps is molecular parachor, V is molar volume, M is molecular weight and ρ is density.

Knowing the surface energy of graphene and the surface tension of the support polymers, one can calculate the interfacial energy between graphene and support polymers and subsequently the adhesion energy by using a relation proposed by Girifalco, Good, and Fowkes [6,7]:

$$\gamma_{GP} = \gamma_G + \gamma_P - 2\sqrt{\gamma_G \cdot \gamma_P} = -E_A \tag{2}$$

$\gamma_P$, $\gamma_G$, $\gamma_{GP}$ and $E_A$ are the surface free energy of graphene (phase 1), the surface free energy of polymer (phase 2) and the interfacial tension between graphene and polymer, and the adhesion energy, respectively (Table S2).

Table S2. Surface energy and binding energy at graphene-polymer interface.

| Sample | Surface energy, mJ/m$^2$ | Adhesion energy, mJ/m$^2$ |
|---|---|---|
| Graphene on Si/SiO$_2$ | 50.64 | NA |
| PMMA | 42.5 | -0.35 |
| ALP | 13.22 | -12.11 |
| ALP:PMMA 1:6 | 38.31 | -0.87 |
| ALP:PMMA 1:4 | 36.64 | -1.13 |
| ALP:PMMA 1:2 | 32.74 | -1.94 |
| ALP:PMMA 1:1 | 27.86 | -3.37 |